\documentclass[fleqn,10pt]{wlscirep}
\usepackage[utf8]{inputenc}
\usepackage[T1]{fontenc}

\usepackage{amssymb}
\usepackage{amsmath}
\usepackage{array}
\usepackage{algorithm}
\usepackage[noend]{algpseudocode}
\usepackage{graphicx}
\graphicspath{ {images/} }
\usepackage{listings}
\usepackage{cite}
\usepackage{float}
\usepackage{comment}

\usepackage[colorinlistoftodos,prependcaption]{todonotes}

\newtheorem{thm}{Theorem}

\newcommand{\pr}{\mathbf{P}}

\title{Geometric explanation of the rich-club phenomenon in complex networks}

\author[1]{Máté Csigi}
\author[1]{Attila Kőrösi}
\author[1]{József Bíró}
\author[1]{Zalán Heszberger}
\author[2]{Yury Malkov}
\author[1,3,*]{András Gulyás}
\affil[1]{MTA-BME Information Systems Research Group, Budapest
  University of Technology and Economics, H-1117 Budapest, Magyar
  tudósok krt. 2, Hungary}
\affil[2]{Federal state budgetary institution of science, Institute of Applied Physics of the Russian
Academy of Sciences, 46 Ul'yanov Street, 603950 Nizhny Novgorod, Russia}
\affil[3]{A. Guly\'as was supported by the Janos Bolyai Fellowship of the Hungarian Academy of Sciences}
\affil[*]{Corresponding author: András Gulyás (gulyas@tmit.bme.hu)}


\begin{abstract}
  The rich club organization (the presence of highly connected hub
  core in a network) influences many structural and functional
  characteristics of networks including topology, the efficiency of
  paths and distribution of load. Despite its major role, the
  literature contains only a very limited set of models capable of
  generating networks with realistic rich club structure.  One
  possible reason is that the rich club organization is a divisive
  property among complex networks which exhibit great diversity, in
  contrast to other metrics (e.g. diameter, clustering or degree
  distribution) which seem to behave very similarly across many
  networks. Here we propose a simple yet powerful geometry-based
  growing model which can generate realistic complex networks with
  high rich club diversity by controlling a single geometric
  parameter. The growing model is validated against the Internet,
  protein-protein interaction, airport and power grid networks.
\end{abstract}
\begin{document}

\flushbottom
\maketitle

\thispagestyle{empty}

\section*{Introduction}

\begin{figure}[ht]
  \center
  \includegraphics[width=0.45\linewidth]{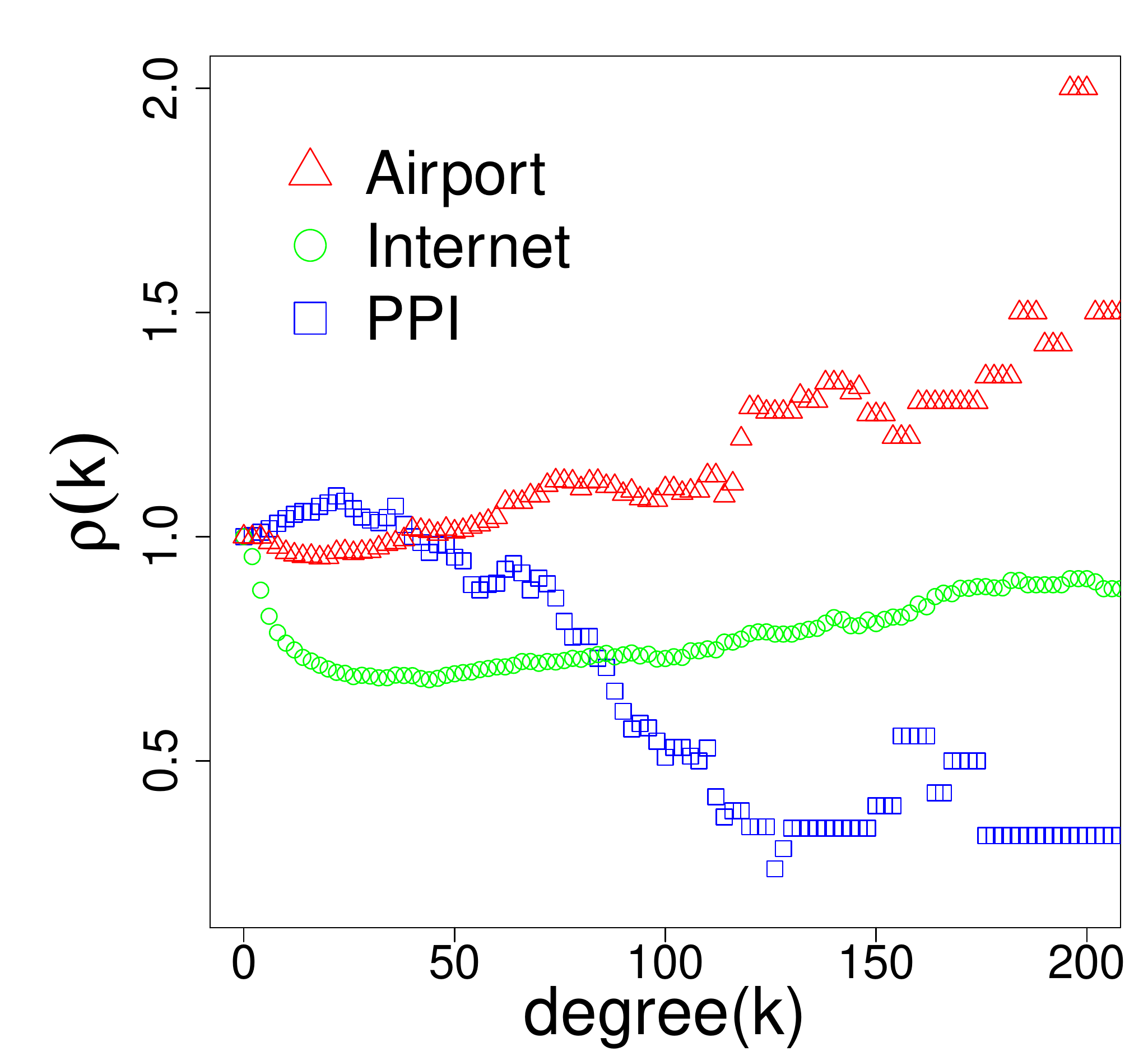}
  \caption{Illustration of the diverse rich club organization in real
    networks. The plot shows the normalized rich club coefficient
    $\rho(k)$ as the function of degree $k$ as:
    $\rho(k)=\frac{\phi(k)}{\phi_{\text{unc}}(k)}$, where $\phi(k)$ is
    the density of the subgraph $G_k$ of the network containing only
    the nodes with $\text{degree}>k$ . The $\phi_{\text{unc}}$ is the
    same for the maximally randomized version of the network
    conserving the degree distribution.  One can see that the subgraph
    of hubs in the airport network possesses about $1.5$ times more
    links than in the randomized version. On the contrary, the hubs in
    the PPI network have less than the half of the connections
    experienced in its randomized counterpart.}
  \label{fig:realRich}
\end{figure}

The rich club organization plays a central role in the structure and
function of networks
\cite{colizza2006detecting,park2013structural,vaquero2013rich,van2013abnormal,ball2014rich,harriger2012rich,van2011rich,zhou2004rich}. Some
networks (e.g. the human brain \cite{van2011rich}, airport networks,
social networks \cite{colizza2006detecting} and the Internet
\cite{zhou2004rich}) have a strong rich club meaning that their hubs
are densely connected to each other. Others (e.g. protein-protein
interaction networks \cite{colizza2006detecting}, the power grid
\cite{mcauley2007rich}) behave quite the contrary as the subgraphs
made out of their hubs are very sparse.  This high variation across
networks is illustrated in Figure~\ref{fig:realRich}, which shows the
normalized rich club coefficient $\rho(k)$ \cite{colizza2006detecting}
as the function of degree $k$ for the airport network, the Internet
and the protein-protein interaction network.
The explanation and reproduction of this great rich club diversity is
highly non trivial. The state-of-the-art models targeting the rich
club organization are based on heavy randomization techniques
\cite{mondragon2012random,mondragon2014network,ma2015rich,xu2010rich},
which shuffle network connections until a given organization structure
is artificially imitated. Although these randomization-based models
are fairly usable, they do not give deeper insight into the mechanisms
causing this diversity during the evolution of the networks.
Consequently, \emph{growing} models capable of incorporating various
rich club networks in a simple and intuitive manner would be useful
towards deeper understanding the underlying evolutionary reasons of
this diversity.

Here we propose a simple geometry-based \emph{growing} model which can
explain the emergence of the rich club variability in real networks by
adjusting a single spatial parameter. Our model is built upon the
real-world observation that in some networks the establishment of very
long connections is not feasible. For example in power grid networks,
the electric current cannot be transferred efficiently (i.e. without
huge losses of energy) over large distances without intermediate
transformations at middle stations
\cite{paris1984present,simpson2016voltage}. Similarly, optical
networks apply signal re-generators for the transmission of light
signals over large distances to be able to sustain the signal-to-noise
ratio \cite{giles1996optical}.  Also in certain social networks,
middlemen as intermediate nodes may play crucial role in enhancing
cooperation between the individuals or groups \cite{borondo2014each}.
Such networks seem to implement an ``artificial'' threshold above
which no direct connections are allowed. Other networks do not have
such inherent thresholds and the length of the edges is only limited
by the ``natural'' geometric boundary of the network. For example in
airport networks we can find very long links, because transferring
passengers over large distances is not an issue with the current
aviation technologies.

In this paper we confine these observations into a simple geometric
growing model, in which we introduce a length threshold for creating
edges. We show that such a growing model can naturally reproduce and
account for the experienced diversity in the rich-club organization of
networks, while keeping other network statistics (diameter, degree
distribution and clustering) intact. The applied geometric
representation of networks is an active and quickly advancing research
direction in network science\cite{cohen2010complex}. There are
numerous studies describing networks as random geometric graphs,
performing some functions
\cite{kleinberg2000navigation,gulyas2015navigable} (e.g. navigation,
information transmission) or structural properties (e.g. small-world,
clustering, modularity)
\cite{papadopoulos2012popularity,malkov2015growing} of networks in a
geometric context, and disclosing some fundamental relations between
topology and hidden metric spaces\cite{serrano2008self}.  A the proper
choice of geometry (e.g. Euclidean, Bolyai-Lobachevskian hyperbolic
geometry or other metric space) can also promote the interpretation of
numerous network processes \cite{boguna2009navigability,
  krioukov2010hyperbolic,allard2017geometric}.

\section*{Results}

\begin{figure}[t]
	\begin{minipage}[b]{0.41\textwidth}
        \centering
		\begin{algorithmic}
		\State Parameters:$N$, $R$, $m$, $T$
		\State $V=\{\}$, $E=\{\}$
		\For{i=0, i<N}
			\State $V=V\bigcup i$
		    \State $i_r=\sqrt{UR^2}$, where $U=\text{random}(0, 1)$
			\State $i_a=\text{random}(0,2\pi)$
		    \For{j in closest $\max(m,|V|)$ nodes to i in V}
				\State $\text{d}_\text{eff} = \text{EuclDist}(i,j)/\sqrt{\text{k}_j}$
				\If{
		  			$\text{d}_\text{eff} < T$} \State $E=E\bigcup (i,j)$
			  	\Else
					\State $V=V\bigcup b^{ij}$,
                    \State $[b^{ij}_r, b^{ij}_a] = \text{midpoint}(\overline{ij})$
		            \State $E=E\bigcup \{(i,b^{ij}),(j,b^{ij})\}$
				\EndIf
			\EndFor
		\EndFor
		\end{algorithmic}
%        \caption{Pseudocode of the model.\label{fig:pseudocode}}
	\end{minipage}
	\begin{minipage}[b]{0.31\textwidth}
          \centering
          \includegraphics[width=0.99\linewidth]{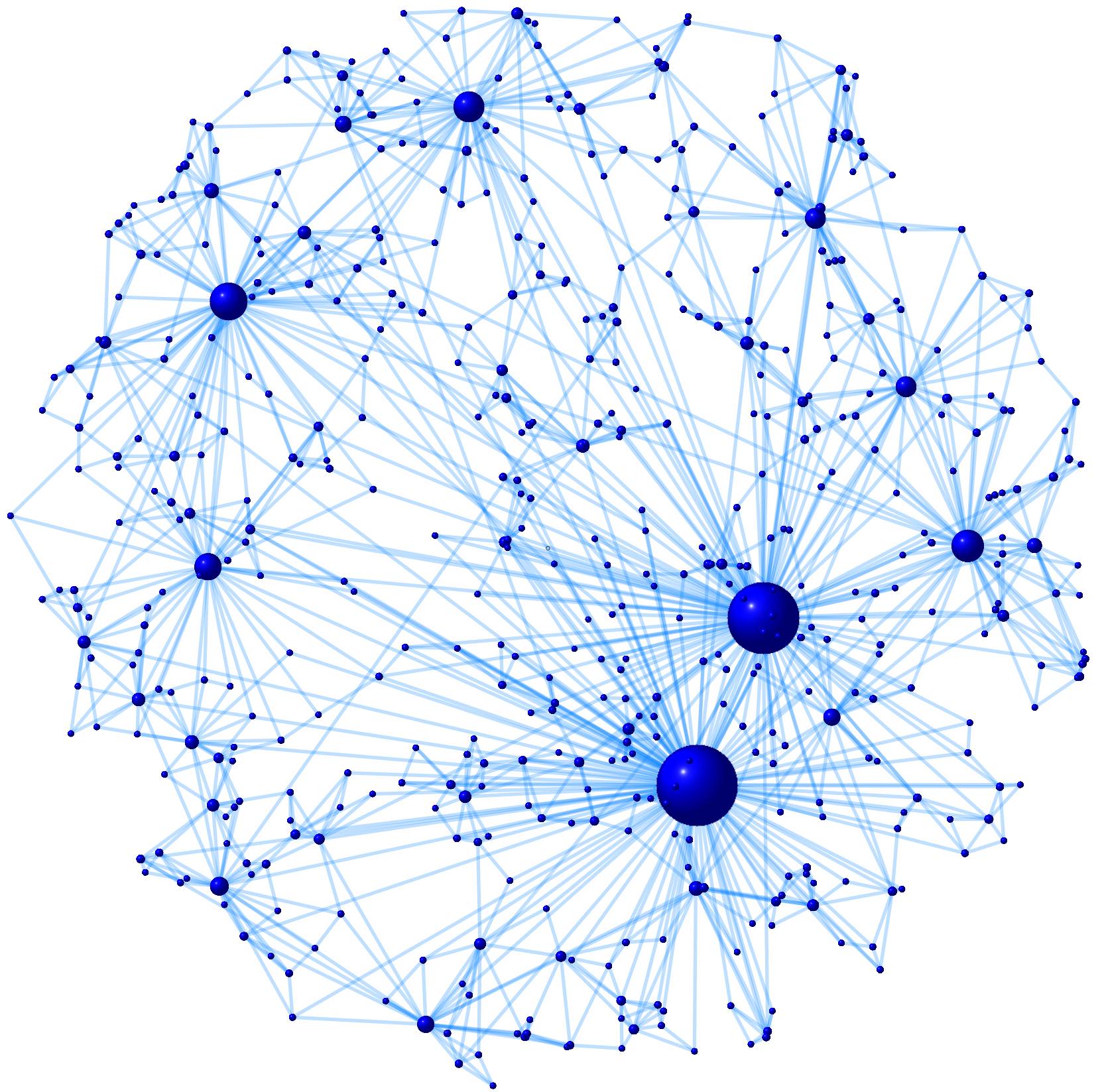}
%          \caption{Sample network generated by the model.\label{fig:network}}
	\end{minipage}
	\begin{minipage}[b]{0.3\textwidth}
          \centering
          \includegraphics[width=0.52\linewidth]{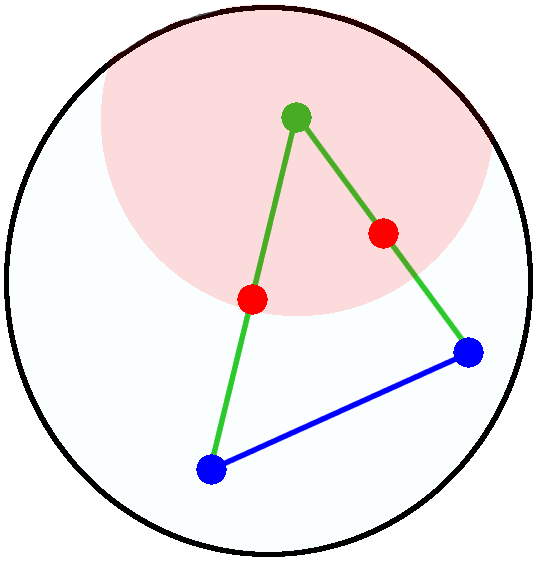}
          \includegraphics[width=0.52\linewidth]{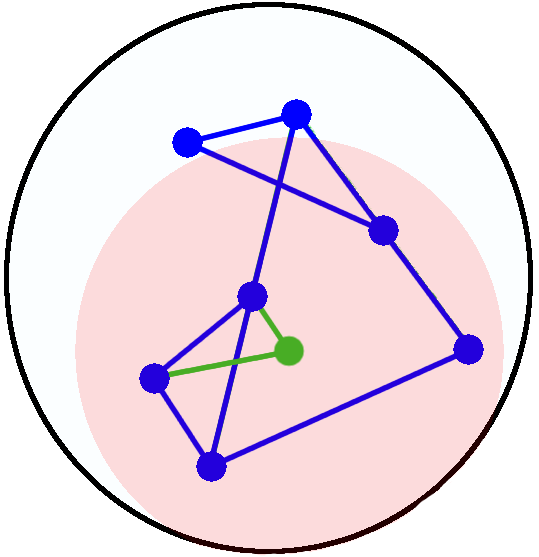}
%          \caption{Sample network generated by the model.\label{fig:network}}
	\end{minipage}
        \put(-505.5,170.5){\large (\textbf{a})}
        \put(-300.5,155.5){\large (\textbf{b})}
        \put(-120.5,155.5){\large (\textbf{c})}
        \put(-120.5,69.5){\large (\textbf{d})}
        \caption{Details of the model. Panel (\textbf{a}) shows the
          exact pseudocode of the generation process. Panel
          (\textbf{b}) plots a sample network generated by the
          model. Panels (\textbf{c}-\textbf{d}) shows time evolution
          of the network for $m=2$. In panel (\textbf{c}) a green node
          is added to a network consisting the two blue nodes
          only. The red circle represents the threshold around the new
          node. As neither of the two blue points reside within the
          red circle two bridge nodes (red) are inserted to the
          midpoints of the edges. Panel (\textbf{d}) shows the same
          network in a later time instant. In this later phase the new
          green node can connect directly (bridge nodes are not
          needed) to the two nearest blue nodes as they are closer
          than $T$.\label{fig:model}}
\end{figure}

In our model $N$ nodes are randomly generated one after another on an
Euclidean 2D $R$-disk with uniformly distributed coordinates. When
adding a new node, it selects the $m$ closest nodes already residing
on the disk (if there are less than $m$ nodes on the disk then it
selects all of them).  The distances between the new node and the old
ones are calculated by the Euclidean distances normalized by a
function of the old node degrees (as in the Growing Homophilic
model\cite{malkov2015growing}).  If this ``effective'' distance
between the new node and a selected one is smaller than the threshold
$T$ then they are directly connected, otherwise a so-called "bridge"
node to the midpoint of the two nodes is established and connects to
both nodes.  The formal description of the network generation process
is performed in panel (\textbf{a}) of Figure~\ref{fig:model}, while
panel (\textbf{b}) shows a small network generated with the model.

Time evolution of the model, as new nodes are inserted into the network
at different stages is shown in Figure~\ref{fig:model}. For the sake of
simplicity, in this illustration the distance normalization by node
degrees is omitted. At the beginning of the
generation process, many bridge nodes are inserted as the distance
between the nodes is typically larger than $T$ (see panel (\textbf{c})
in Figure~\ref{fig:model}). As the network grows, the average node density and
degrees increases, so the typical normalized distance between the
nodes will fall below $T$ and no more bridge nodes are added (panel
(\textbf{d}) in Figure~\ref{fig:model}). From this stage the model
falls back to the growing homophilic model analyzed in
\cite{malkov2015growing}. Setting $T$ to a very large value
(e.g. $T>2R$) completely recovers the model in
\cite{malkov2015growing} because bridge nodes are never inserted to
the graph.  We show, that by varying $T$, the model generates complex
networks with diverse rich-club organization, while having scale-free
degree distribution, small diameter and large clustering.  In the
remaining of the paper we will use the settings summarized in
Table~\ref{tab:simSettings} in our analytical and simulation results.

\begin{table}
  \begin{center}
    \begin{tabular}{ | c | c | c | c | c |}
      \hline
      Network & $N$ & $m$ & $R$ & $T$  \\ \hline
      Generated $T=12$ & 5000 & 3 & 50 & 12 \\ \hline
      Generated $T=30$ & 5000 & 3 & 50 & 30 \\ \hline
      Generated $T=100$ & 5000 & 3 & 50 & 100 \\ \hline
    \end{tabular}
  \end{center}
  \caption{Simulation settings.}
  \label{tab:simSettings}
\end{table}

\subsection*{Number of bridge nodes}

First, we show that the total number of bridge nodes quickly converges
to a relatively small value compared to the reasonable network size
($N$) during the generation of the graph, and this value is
independent of the graph size. To support this observation we give a
recursive estimation of the expected number of new bridge nodes
generated at each step of the model, and based on this recursion a
mathematical expression is given to the limit of the expected total
number of bridge nodes (see methods for more details Methods).
By analyzing the recursion one can show that the expected number of bridge nodes at step $N$ denoted by $b_N$ can approximately be expressed in the form
\begin{equation}
b_N \approx \exp(-f_1 N + f_2 \log N + f_3 ) \ .
\end{equation}
where the functions $f_1$, $f_2$ and $f_3$ may depend on $R,T,m$ but
are independent from $N$. From this it immediately follows, that for the total number of bridge nodes $B_N$
\begin{equation}
B_N \approx \int_{x=1}^N \exp(-f_1 x + f_2 \log x + f_3 ) {\rm d} x \rightarrow \exp(f_3) { E}_{-f_2}(f_1) \ {\rm as} \ N \rightarrow \infty
\end{equation}
where ${E}$ is the exponential integral function. The vanishing term
during the convergence in $B_N$ is $N^{1+f_2} E_{-f_2} (f_1 N)$ and
also approximately exponential.

\begin{figure}[ht]
  \begin{minipage}[t]{.49\textwidth}
    \center
  \includegraphics[width=0.9\linewidth]{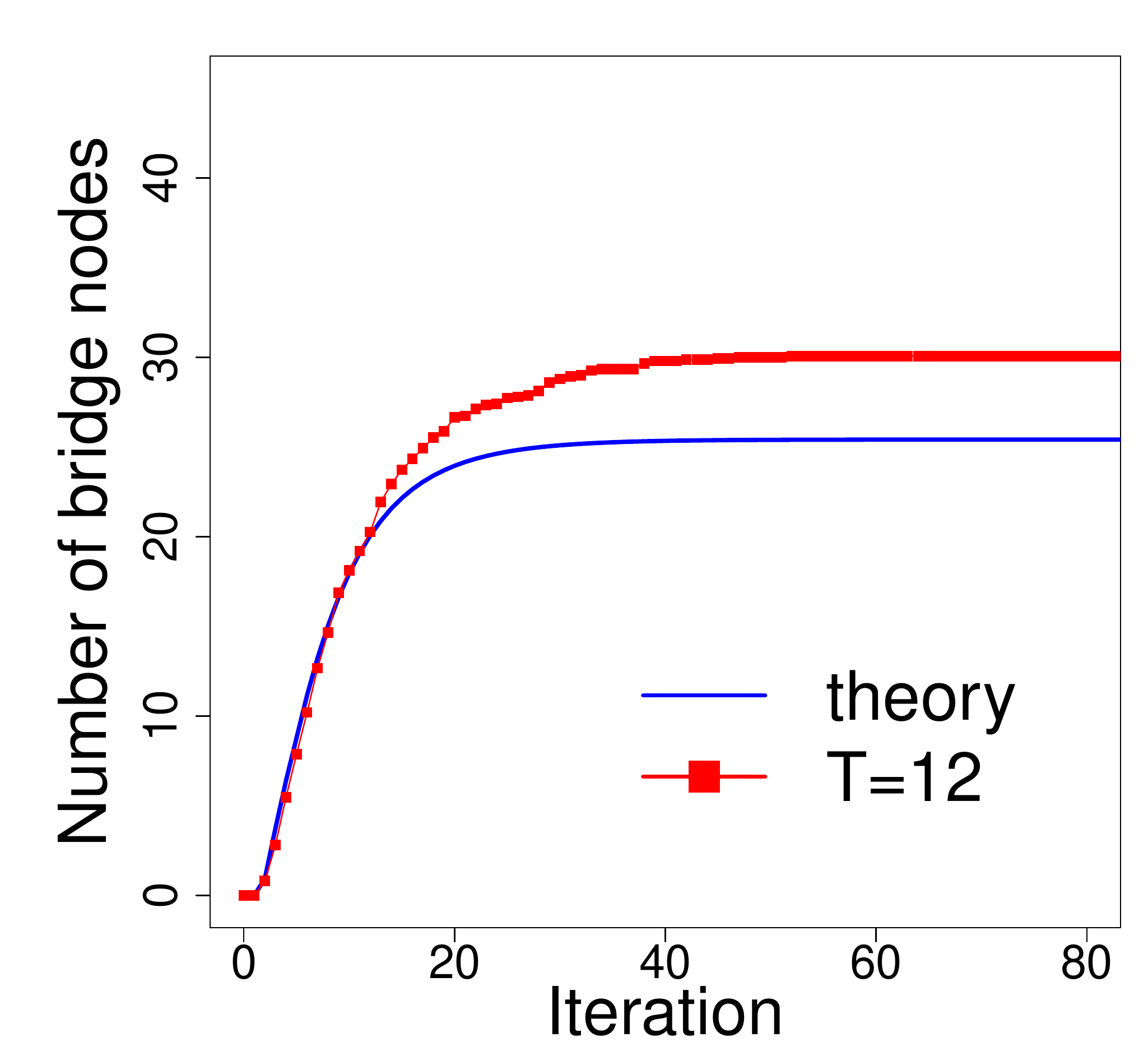}
  \caption{The recursively estimated number of bridge nodes in each
    iteration is plotted with the simulated values. The $T=12$ line
    shows the average of $15$ simulation runs.}
  \label{fig:bridgeTime}
  \end{minipage}
  \hfill
  \begin{minipage}[t]{.49\textwidth}
    \includegraphics[width=0.9\linewidth]{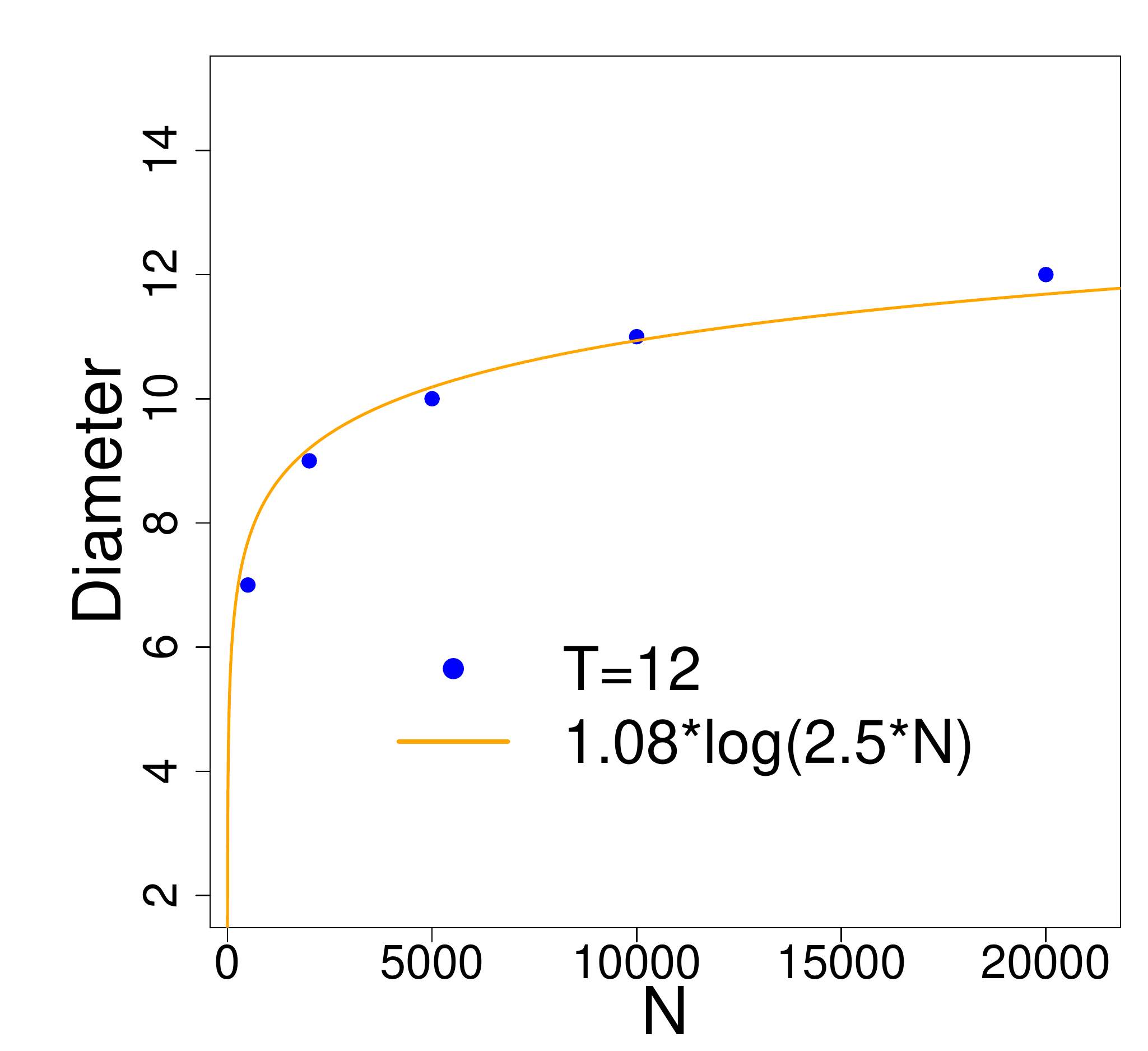}
    \caption{Small world property of $T=12$ networks. The diameter of
      the network grows logarithmically with the number of nodes.}
    \label{fig:smallWorld}
  \end{minipage}
\end{figure}

In Figure~\ref{fig:bridgeTime} the expected total number of bridge
nodes ($B_N$) calculated by recursion (\ref{equ:recursion}) is plotted
in each iteration together with the simulation result for the same
parameters. The two plots readily justify that $B_N$ has a
characteristic flat after certain iterations, which means that $B_N$
converges to a finite fixed value during the graph generation
process. This also illustrates that for sufficiently large network the
total number of bridge nodes is negligible comparing to the network
size. Furthermore, according to statistical tests the overall
distribution of the nodes on the $R$-disk is apparently not affected
by the bridge nodes, and still can be treated as uniform.

% \begin{figure}[ht]
%   \center
%   \includegraphics[width=0.4\linewidth]{bridgeTime.pdf}
%   \caption{The recursively estimated number of bridge nodes in each
%     iteration is plotted with the simulated values. The $T=12$ line
%     shows the average of $15$ simulation runs.}
%   \label{fig:bridgeTime}
% \end{figure}

\subsection*{Diameter, clustering and degree distribution}

The diameter of all three generated networks (see
Table~\ref{tab:simSettings}) is around $9$-$10$, similar to the real
networks (Table~\ref{tab:netProperties}). Figure~\ref{fig:smallWorld}
shows that the diameter of the $T=12$ networks is an approximate
logarithmic function of the network size, which confirms the
small-world property. Also the generated networks have high clustering
coefficients with values very close to that of real networks.
Finally, Table~\ref{tab:netProperties} confirms that the clustering
coefficient is insensitive to the threshold parameter.
\begin{table}
  \begin{center}
    \begin{tabular}{ | c | c | c | c | c | c |}
      \hline
      Network & N & Edges & Diameter & Avg. dist. & Clustering coefficient \\ \hline
      Generated $T=12$ & 5030 & 15024 & 10 & 4.54 & 0.66\\ \hline
      Generated $T=30$ & 5004 & 14998 & 9 & 4.51 & 0.67\\ \hline
      Generated $T=100$ & 5000 & 14994 & 10 & 4.46 & 0.69\\ \hline
      PPI network & 5084 & 22148 & 10 & 3.98 & 0.12\\ \hline
      Airport network & 2845 & 10409 & 10 & 3.75 & 0.59\\ \hline
      Internet & 23748 & 58414 & 10 & 3.52  & 0.61\\ \hline
    \end{tabular}
  \end{center}
  \caption{Basic topological properties of real and generated networks.}
  \label{tab:netProperties}
\end{table}
Now we show that the generated networks has scale-free degree
distribution independently of $T$.

\begin{thm}
  The networks produced by the model have scale-free degree
  distribution with $\gamma=3$ when $N\to\infty$.
\end{thm}

\noindent \textbf{Proof:} Suppose we compute the effective distance as
${{d}_{\text{eff}}}=\frac{{{d}_{Euc}}}{\sqrt{k}}$.  At each insertion
step the algorithm connects a new element to exactly $m$ neighbors
that globally minimize the normalized distance.  To infer the degree
distribution of the neighbor elements, we temporary fix the distance
to the $m+1$-th nearest neighbor $d_{\text{eff}}^{m+1}$ and randomly
shuffle positions of the $m$ neighbor nodes under the condition that
they all remain the $m$ nearest neighbors with respect to the new
element (i.e. having effective distance to the new element less than
$d_{\text{eff}}^{m+1}$). \\
For every possible value of the neighbor degree $k$, possible element
positions are bounded in the initial Euclidean space by a radius
${{r}_{\text{Euc}}}=d_{\text{eff}}^{m+1}\sqrt{k}$. Since the nodes are
distributed uniformly in the Euclidean space, the probability of
having an element with degree k proportional to the
$r_{\text{Euc}}$-ball volume. Thus under fixed $d_{\text{eff}}^{m+1}$
the overall probability of connecting to an element with degree $k$ is
proportional to $(k)$. \\
The probability inferred for a fixed value of $d_{\text{eff}}^{m+1}$
does not depend on either the value of $d_{\text{eff}}^{m+1}$, or the
positions of the nodes that are not the closest neighbors of the
inserted elements, so that is true for every possible positions of the
elements in the Euclidean space and overall probability of connection
to a node is proportional to its degree $(k)$. This means that new
nodes connect to the old ones with probability proportional to $k$,
which is equivalent to the Barabasi-Albert model
\cite{barabasi1999emergence}, proved to produce scale-free networks
with $\gamma =3$.
\begin{comment}
In the following we consider that modificated model, which connects the new node $u$ to nodes  are inside a ball with $tr_{eff}$ effective threshold radial instead of the $m$ closest ones (we will choose $tr_{eff}$ such that, the expected number of the new edge would be $m$). Probability that $u$ will be connected with node $v$ if the degree of $v$ is $k$:
\begin{multline*}
\pr(d_{eff}(u,v)<tr_{eff})=\pr\left(\frac{d_{\mathrm{Euc}}(u,v)}{f(k)}<tr_{eff}\right)=\pr\left(d_{\mathrm{Euc}}(u,v)<tr_{eff} f(k)\right)\\
\approx\frac{(tr_{eff} f(k))^D}{R^D}\sim f(k)^D
\end{multline*}
If $f(k)=\sqrt[D]{k}$, then the connection probability is proportional to $k$, which is statistically equivalent to Barabasi-Albert model, and provides scale-free degree distribution.
To determine the suitable $tr_{eff}$ we have to calculate the expected degree of $u$:
\begin{multline*}
\mathbf{E}(deg(u))=\sum_{v}\pr(d_{eff}(u,v)<tr_{eff})\approx\sum_v \frac{(tr_{eff} f(k))^D}{R^D} = \frac{tr_{eff}^D}{R^D}\sum_vf(k)^D
\end{multline*}
In the case of $D=2$ and $f(k)=\sqrt{k}$, $N=\delta R^2 \pi$, where $\delta$ is the density of the nodes on the disk, moreover the sum of the degrees of the nodes is closely $2 N m$, because its twice of the number of edges.
Hence, if the expected degree is $m$:
\begin{multline*}
m=\frac{tr_{eff}^2}{R^2}\sum_vk=\frac{tr_{eff}^2}{\tfrac{N}{\delta\pi}}2 N m = m \frac{tr_{eff}^2}{\tfrac{1}{2\delta\pi}}
\end{multline*}
So, $tr_{eff}\approx frac1{\sqrt{2\delta\pi}} = \tfrac{R}{\sqrt{2N}}$
\end{comment}

\begin{figure}[H]
  \begin{minipage}[t]{.49\textwidth}
    \center
    \includegraphics[width=0.9\linewidth]{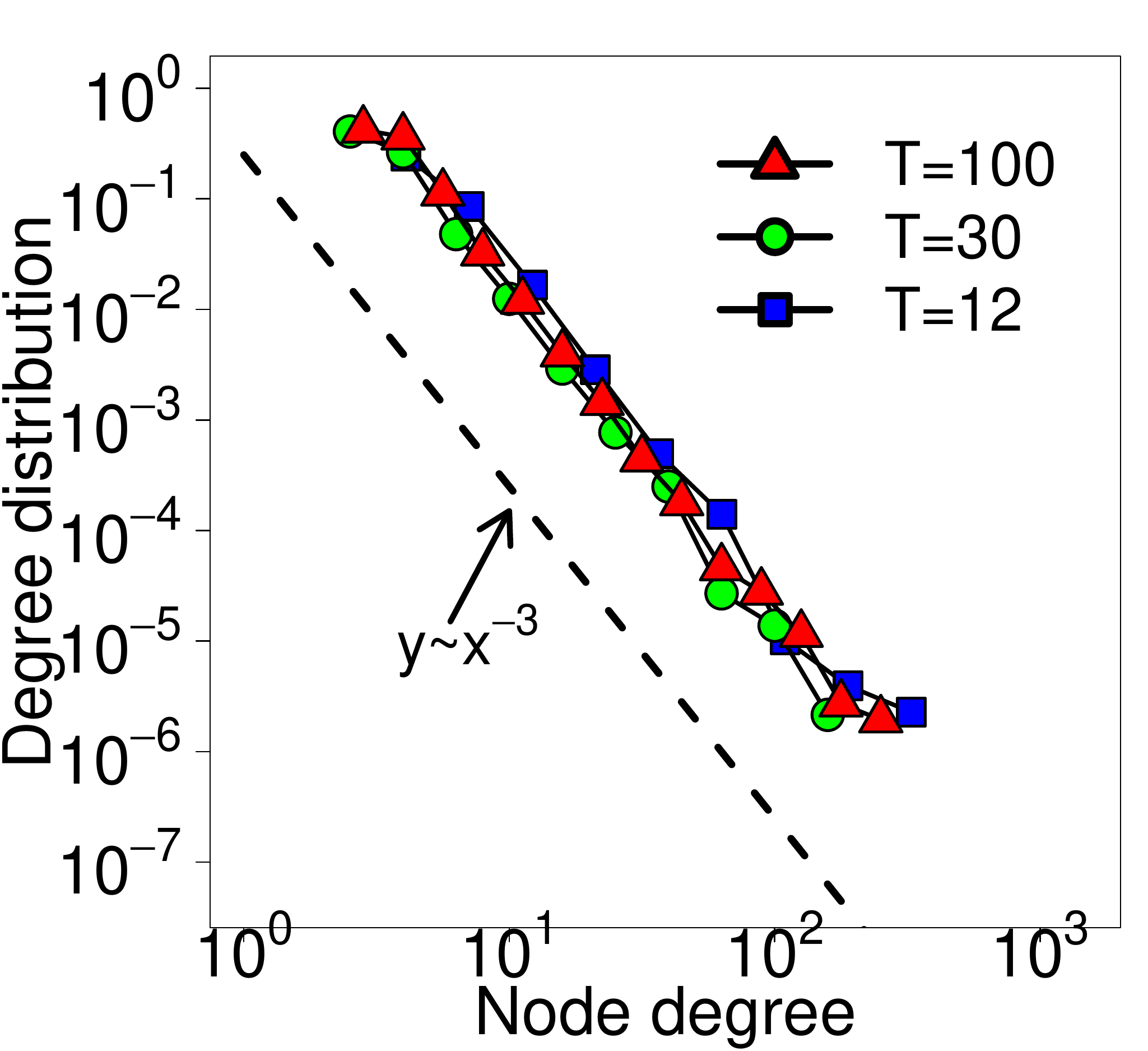}
    \caption{The degree distributions of our simulated networks. The
      plot confirms that the choice of $T$ does not effect the degree
      distribution, which is a power-law with $\gamma=3$.}
    \label{fig:degreeDist}
  \end{minipage}
  \hfill
  \begin{minipage}[t]{.49\textwidth}
    \includegraphics[width=0.9\linewidth]{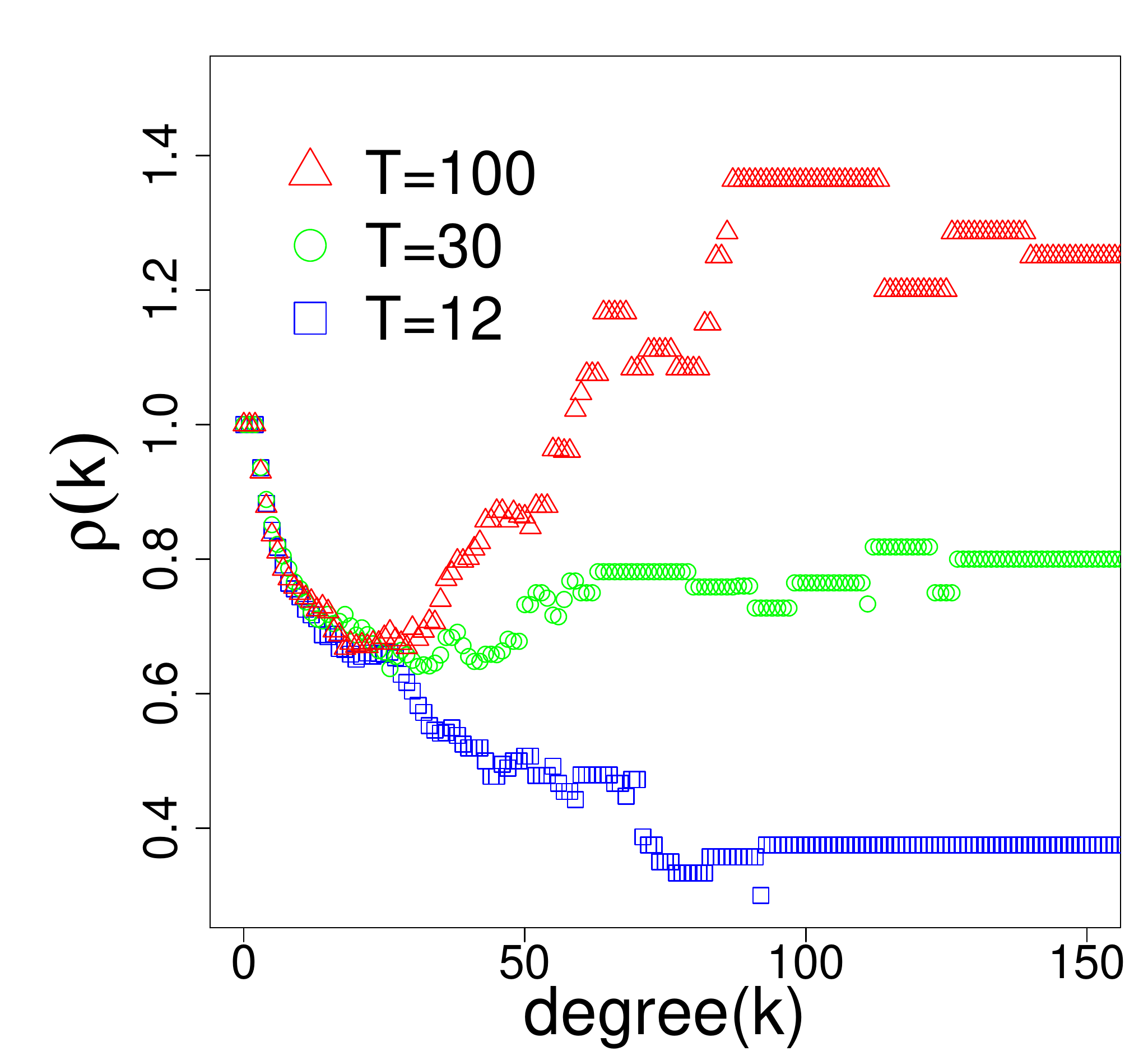}
    \caption{Rich-club organization in networks generated by the model
      with different settings of $T$. The plot readily confirms that our
      model is able to generate rich-club diversity by simply adjusting
      $T$. We note for the remarkable similarity with the same plot for
      real networks in Figure~\ref{fig:realRich}.}
    \label{fig:generatedRich}
  \end{minipage}
\end{figure}

The Figure~\ref{fig:degreeDist} shows the degree distributions of three
networks generated with our model with various values of $T$. The plot
readily confirms that the degree distributions are indeed scale-free
with $\gamma=3$ independently of $T$.

\subsection*{Rich-club coefficient}

Although the insertion of bridge nodes keeps degree distribution,
clustering and diameter intact, the simulation results plotted in
Figure~\ref{fig:generatedRich} clearly show that the graphs generated
by the model differ greatly in their rich-club organization depending
on $T$. Setting $T$ to the diameter of the $R$-disk ($T=100$, red
triangles in Figure~\ref{fig:generatedRich}), the model does not limit
the lengths of the edges artificially, so the only limiting factor is
the natural geometry of the disk itself. In this case we obtain a
network with a strong rich-club, similarly to the airport
network. Conversely, adjusting $T$ to $12$, the model will create only
edges having $d_{\text{eff}}<12$. This is a strong ``artificial''
limitation for the edge lengths imposed by the generation process. As
a result, the model yields a network with no rich-club ($T=12$, blue
squares in Figure~\ref{fig:generatedRich}), likewise the PPI network.
We note the appealing similarity between
Figure~\ref{fig:generatedRich} and Figure~\ref{fig:realRich}, showing
the rich-club diversity in real networks.

\section*{Discussion}

\begin{figure}[ht]
\begin{center}
  \includegraphics[width=0.49\linewidth]{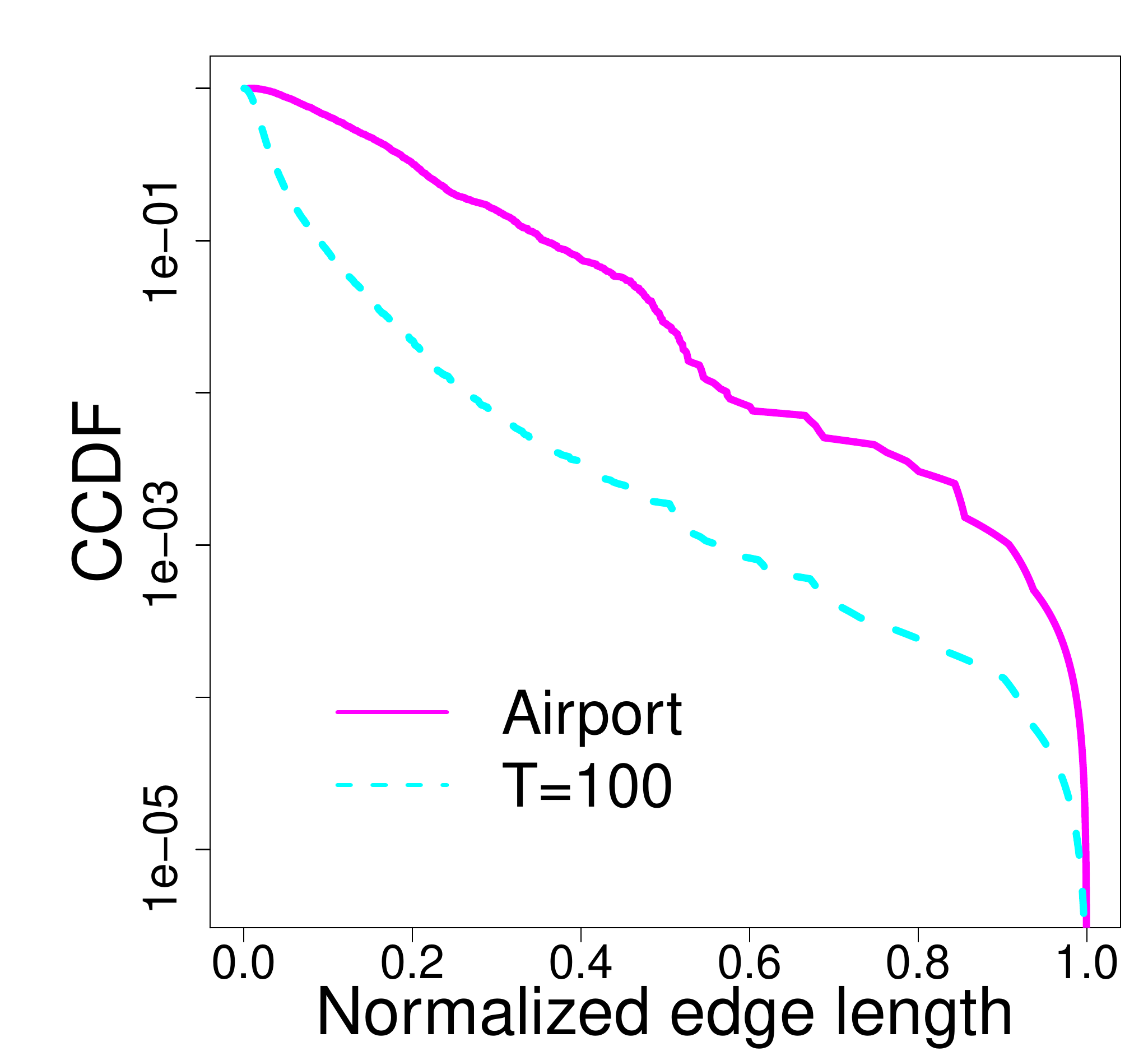}
  \includegraphics[width=0.49\linewidth]{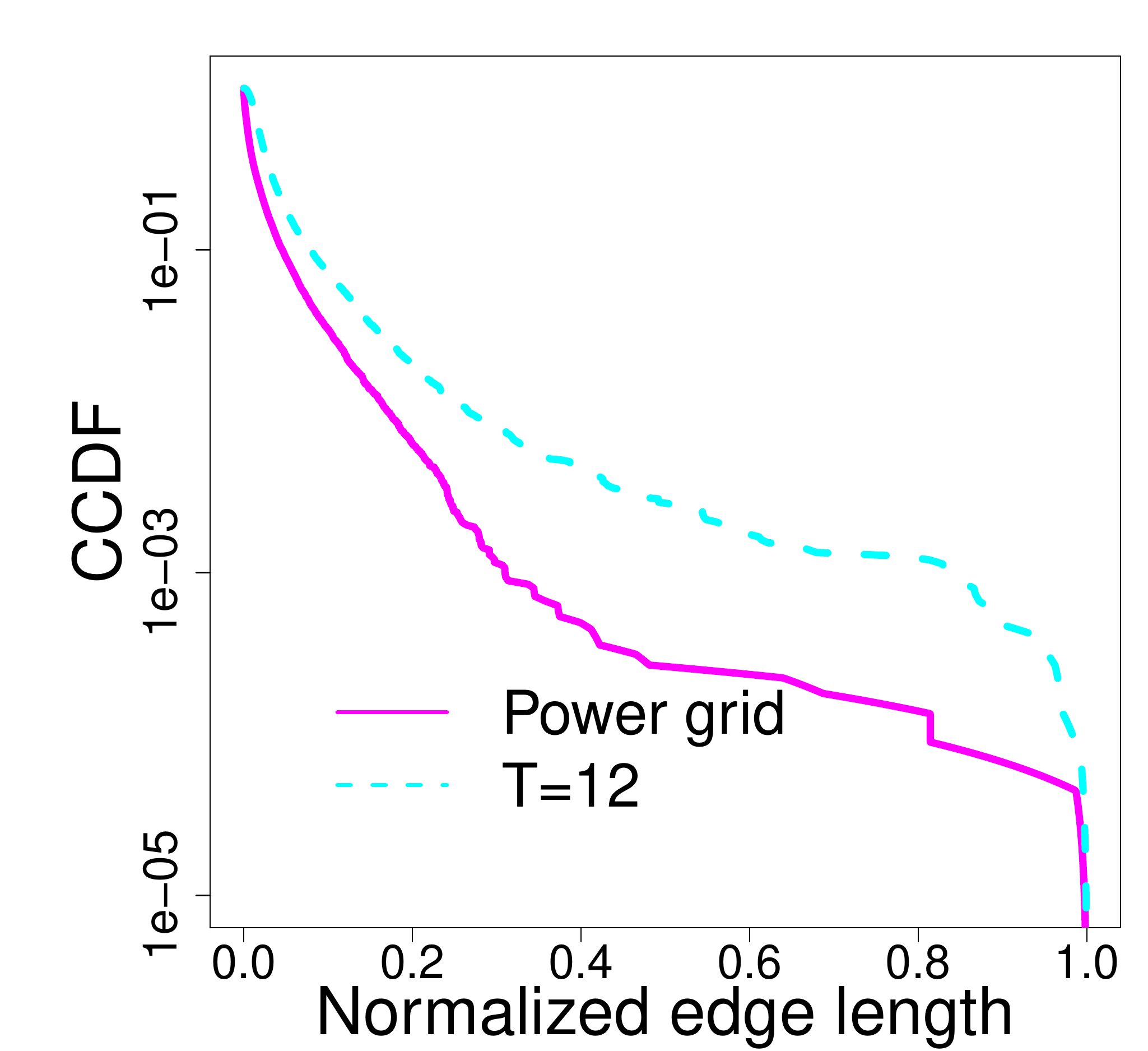}
  \put(-470.5,200.5){\large (\textbf{a})}
  \put(-220.5,200.5){\large (\textbf{b})}
  \put(-330.5,115.5){\rotatebox{-31}{\large No plateau}}
  \put(-68.5,90.5){\large Plateau}
  \caption{Edge length distribution in networks with different rich
    club organization. While panel (\textbf{a}) shows a continuously
    decreasing length distribution in case of rich-club networks,
    there is a clearly visible plateau before the cutoff of the CCDF
    function in networks having no rich-clubs (panel (\textbf{b})).}
  \label{fig:edgeLengthDistribution}
\end{center}
\end{figure}

An intriguing question could be whether our model captures something
fundamental from the growth processes of real networks, or exhibit
similar rich-club diversity simply by chance. For answering this
question we have performed the CCDF's (Complementary Cumulative
Distribution Function) of the normalized edge length distribution in a
rich-club (airports with flights in the US) and a non rich-club
network (the North American Power Grid) together with the networks
generated with our model in
Figure~\ref{fig:edgeLengthDistribution}. Panel (\textbf{a}) shows
continuously significant (on all length scale) decrease of edge length
distributions before the final ``natural'' cutoff for the airport and
the $T=100$ networks caused by the geometry of the continent and the
$R$-disk respectively. On panel (\textbf{b}) however we can observe a
clearly visible plateau before the cutoff of the edge lengths in the
power grid network. This means that edge lengths are much denser near
the cutoff, which in this case is rather ``artificial'' and caused by
the growth process of the network and not the underlying geometry. Our
model produces a very similar edge length distribution for the setting
$T=12$. These results hint that networks having no rich-clubs use a
very similar limiting for the length of the connections as our model
do. As a consequence, this length-limiting phenomenon can also account
for the emergence of the observed diverse rich-club organization in
real networks.

These two examples also underline that our method is parsimonious in a
sense that the rich club organization can be tuned by only a single
geometric threshold parameter in a growing homophilic model. We think
the results presented in this paper are strong indications that the
rich club diversity can be placed at all on a growing/evolutionary
perspective, and provide deeper insight into the mechanisms
resulting certain rich club behavior during the growth of networks.

\section*{Methods}

\subsection*{Recursive estimation of the number of bridge nodes}

Let $A(r,T,R)$ be the area of the intersection of an $r-$centered disk
with radius $T$ and the $R-$disk, and let $p(r,T,R)$ be the fraction
of $A(r,T,R)$ and the area of the $R-$disk, i.e.
$p(r,T,R)=\frac{A(r,T,R)}{R^2 \pi}$.  Further, let us assume that
there are already $j$ nodes in the network. The $(j+1)^{\rm th}$
randomly generated node will connect to the $m$ nearest neighbors. For
calculating the necessary bridge nodes in this step, the task is to
determine what are the nodes among the $m$ nearest neighbors which are
farther than $T$. To ease the computation, the degrees of the
neighbors are substituted by their expectation values (denoted by
$\bar k_j$ and to be determined later) subject to the whole network at
this stage. Since the effective distance is computed as the Euclidean
distance divided by $\sqrt{\bar k_j}$, it is approximately equivalent
to investigate the expected number of points among the $m$ nearest
ones being outside of the $(j+1)^{\rm th}$ node $T \sqrt{\bar k_j}$ -
radius vicinity. This will be equal to the expected number of newly
inserted bridge nodes at this step. Denote the radial coordinate of
the $(j+1)^{\rm th}$ node by $r$ and assume that the previously
generated random points and established bridge nodes are still evenly
distributed on the $R-$disk. With this, the probability that
$i, 0 \leq i \leq j$ nodes among the $j$ ones are closer to the
$(j+1)^{\rm th}$ node than $T \sqrt{\bar k_j}$ is
$$
\binom{j}{i} p(r,T \sqrt{\bar k_j},R)^{i} (1-p(r,T \sqrt{\bar k_j},R))^{j-i}
$$
and hence the expected number of necessary bridge nodes at this step is
$$
\sum_{i=0}^{\min(m-1,j-1)}\min(m-i,j-i) \binom{j}{i} p(r,T \sqrt{\bar k_j},R)^{i} (1-p(r,T \sqrt{\bar k_j},R))^{j-i} := \beta_j(r,T \sqrt{\bar k_j},R,m)
$$
Note, that this is still a conditional expectation value which is to
be de-conditioned by the density of the radial coordinate $r$. Towards
the de-conditioning, first the function $A(r,T,R)$ is to be
determined. Clearly, $A(r,T,R) = T^2 \pi$ if $r \leq R-T $, i.e. there
is no intersection of the two disks. Otherwise, if $r \geq R-T $ then
by using straightforward geometrical calculations
\begin{equation}
A(r,T,R) = \alpha R^2 +\gamma T^2-2 \sqrt{s (s-R) (s-T)(s-r)}
\end{equation}
where
\begin{equation}
\alpha = \arccos \frac{R^2+r^2-T^2}{2 r R} \ , \ \gamma= \arccos \frac{r^2+T^2-R^2}{2 r T} \ , \
s=\frac{R+T+r}{2} \ .
\end{equation}
Now, the de-conditioning is possible with
$p(r,T,R)=A(r,T,R)/(R^2 \pi)$ and the density of the radial coordinate
$r$, which is $\frac{2 r}{R^2}$. Further, let $j(N)=N+b_1+b_2+\ldots+b_N$ where
$N$ is the randomly generated points and $b_l \ l=1,\ldots,N$ is the
expected number of bridge nodes established after the $l^{\rm th}$
random node. For completing the recursive estimation, the expected
degree should also be expressed upon the $l^{\rm th}$ random node
generation. This is
\begin{equation}
\bar k_l = \frac{2 (l m - \frac{m(m+1)}{2}+2 \sum_{i=1}^l b_i)}{l+\sum_{i=1}^{l} b_i} \ , \ {\rm for} \  l > m \ , \ {\rm else} \ \bar k_{l} =\frac{2 (\frac{l(l-1)}{2}+2 \sum_{i=1}^{l} b_i)}{l+\sum_{i=1}^{l} b_i} \ .
\end{equation}
For $l=1$ let $b_1=0$, $\bar k_1=0$ and let $B_N = \sum_{i=1}^N b_i$.
The main recursion can now be expressed as
% \begin{equation} \label{equ:recursion}
% \! \! \! \! \! \! \! \! \! \! \! \! b_{N+1} = \int_{r=0}^R \sum_{i=0}^{\min(m-1,N+B_N-1)} \! \! \! \! \! \! \! \!  \min(m-i,N+B_N-i) \binom{N+B_N}{i} p(r,T \sqrt{\bar k_N},R)^{i} (1-p(r,T \sqrt{\bar k_N},R))^{N+B_N-i}  \frac{2 r}{R^2} {\rm d} r \ .
% \end{equation}
\begin{equation} \label{equ:recursion}
\! \! \! \! \! \! \! \! \! \! \! \! b_{N+1} = \int_{r=0}^R
\beta_{j(N)}(r,T \sqrt{\bar k_{j(N)}},R,m) \frac{2 r}{R^2} {\rm d} r \ .
\end{equation}

\subsection*{Data Availability}

The data that support the findings of this study are available from public data repositories. In particular, the topology of the AS level Internet has been downloaded from CAIDA (Center for Applied Internet Data Analysis, \url{www.caida.org}). We have downloaded the airport network from the OpenFlights database (\url{www.openflights.org}). We used the DIP \cite{xenarios2002dip} database as a source for the protein-protein interaction network of the S. cerevisiae. Finally, the map of the north american power grid has been downloaded from \cite{wiegmans_2016_47317}.

\bibliography{ref.bib}

\begin{thebibliography}{10}
\expandafter\ifx\csname url\endcsname\relax
  \def\url#1{\texttt{#1}}\fi
\expandafter\ifx\csname urlprefix\endcsname\relax\def\urlprefix{URL }\fi
\expandafter\ifx\csname doiprefix\endcsname\relax\def\doiprefix{DOI }\fi
\providecommand{\bibinfo}[2]{#2}
\providecommand{\eprint}[2][]{\url{#2}}

\bibitem{colizza2006detecting}
\bibinfo{author}{Colizza, V.}, \bibinfo{author}{Flammini, A.},
  \bibinfo{author}{Serrano, M.~A.} \& \bibinfo{author}{Vespignani, A.}
\newblock \bibinfo{title}{Detecting rich-club ordering in complex networks}.
\newblock \emph{\bibinfo{journal}{Nature physics}}
  \textbf{\bibinfo{volume}{2}}, \bibinfo{pages}{110--115}
  (\bibinfo{year}{2006}).

\bibitem{park2013structural}
\bibinfo{author}{Park, H.-J.} \& \bibinfo{author}{Friston, K.}
\newblock \bibinfo{title}{Structural and functional brain networks: from
  connections to cognition}.
\newblock \emph{\bibinfo{journal}{Science}} \textbf{\bibinfo{volume}{342}},
  \bibinfo{pages}{1238411} (\bibinfo{year}{2013}).

\bibitem{vaquero2013rich}
\bibinfo{author}{Vaquero, L.~M.} \& \bibinfo{author}{Cebrian, M.}
\newblock \bibinfo{title}{The rich club phenomenon in the classroom}.
\newblock \emph{\bibinfo{journal}{Scientific reports}}
  \textbf{\bibinfo{volume}{3}} (\bibinfo{year}{2013}).

\bibitem{van2013abnormal}
\bibinfo{author}{van~den Heuvel, M.~P.} \emph{et~al.}
\newblock \bibinfo{title}{Abnormal rich club organization and functional brain
  dynamics in schizophrenia}.
\newblock \emph{\bibinfo{journal}{JAMA psychiatry}}
  \textbf{\bibinfo{volume}{70}}, \bibinfo{pages}{783--792}
  (\bibinfo{year}{2013}).

\bibitem{ball2014rich}
\bibinfo{author}{Ball, G.} \emph{et~al.}
\newblock \bibinfo{title}{Rich-club organization of the newborn human brain}.
\newblock \emph{\bibinfo{journal}{Proceedings of the National Academy of
  Sciences}} \textbf{\bibinfo{volume}{111}}, \bibinfo{pages}{7456--7461}
  (\bibinfo{year}{2014}).

\bibitem{harriger2012rich}
\bibinfo{author}{Harriger, L.}, \bibinfo{author}{Van Den~Heuvel, M.~P.} \&
  \bibinfo{author}{Sporns, O.}
\newblock \bibinfo{title}{Rich club organization of macaque cerebral cortex and
  its role in network communication}.
\newblock \emph{\bibinfo{journal}{PloS one}} \textbf{\bibinfo{volume}{7}},
  \bibinfo{pages}{e46497} (\bibinfo{year}{2012}).

\bibitem{van2011rich}
\bibinfo{author}{Van Den~Heuvel, M.~P.} \& \bibinfo{author}{Sporns, O.}
\newblock \bibinfo{title}{Rich-club organization of the human connectome}.
\newblock \emph{\bibinfo{journal}{The Journal of neuroscience}}
  \textbf{\bibinfo{volume}{31}}, \bibinfo{pages}{15775--15786}
  (\bibinfo{year}{2011}).

\bibitem{zhou2004rich}
\bibinfo{author}{Zhou, S.} \& \bibinfo{author}{Mondrag{\'o}n, R.~J.}
\newblock \bibinfo{title}{The rich-club phenomenon in the internet topology}.
\newblock \emph{\bibinfo{journal}{IEEE Communications Letters}}
  \textbf{\bibinfo{volume}{8}}, \bibinfo{pages}{180--182}
  (\bibinfo{year}{2004}).

\bibitem{mcauley2007rich}
\bibinfo{author}{McAuley, J.~J.}, \bibinfo{author}{da~Fontoura~Costa, L.} \&
  \bibinfo{author}{Caetano, T.~S.}
\newblock \bibinfo{title}{Rich-club phenomenon across complex network
  hierarchies}.
\newblock \emph{\bibinfo{journal}{Applied Physics Letters}}
  \textbf{\bibinfo{volume}{91}}, \bibinfo{pages}{084103}
  (\bibinfo{year}{2007}).

\bibitem{mondragon2012random}
\bibinfo{author}{Mondrag{\'o}n, R.~J.} \& \bibinfo{author}{Zhou, S.}
\newblock \bibinfo{title}{Random networks with given rich-club coefficient}.
\newblock \emph{\bibinfo{journal}{The European Physical Journal B}}
  \textbf{\bibinfo{volume}{85}}, \bibinfo{pages}{1--6} (\bibinfo{year}{2012}).

\bibitem{mondragon2014network}
\bibinfo{author}{Mondrag{\'o}n, R.~J.}
\newblock \bibinfo{title}{Network null-model based on maximal entropy and the
  rich-club}.
\newblock \emph{\bibinfo{journal}{Journal of Complex Networks}}
  \textbf{\bibinfo{volume}{2}}, \bibinfo{pages}{288--298}
  (\bibinfo{year}{2014}).

\bibitem{ma2015rich}
\bibinfo{author}{Ma, A.} \& \bibinfo{author}{Mondrag{\'o}n, R.~J.}
\newblock \bibinfo{title}{Rich-cores in networks}.
\newblock \emph{\bibinfo{journal}{PloS one}} \textbf{\bibinfo{volume}{10}},
  \bibinfo{pages}{e0119678} (\bibinfo{year}{2015}).

\bibitem{xu2010rich}
\bibinfo{author}{Xu, X.-K.}, \bibinfo{author}{Zhang, J.} \&
  \bibinfo{author}{Small, M.}
\newblock \bibinfo{title}{Rich-club connectivity dominates assortativity and
  transitivity of complex networks}.
\newblock \emph{\bibinfo{journal}{Physical Review E}}
  \textbf{\bibinfo{volume}{82}}, \bibinfo{pages}{046117}
  (\bibinfo{year}{2010}).

\bibitem{paris1984present}
\bibinfo{author}{Paris, L.} \emph{et~al.}
\newblock \bibinfo{title}{Present limits of very long distance transmission
  systems}.
\newblock \emph{\bibinfo{journal}{Global Energy Network Institute}}
  (\bibinfo{year}{1984}).

\bibitem{simpson2016voltage}
\bibinfo{author}{Simpson-Porco, J.~W.}, \bibinfo{author}{D{\"o}rfler, F.} \&
  \bibinfo{author}{Bullo, F.}
\newblock \bibinfo{title}{Voltage collapse in complex power grids}.
\newblock \emph{\bibinfo{journal}{Nature communications}}
  \textbf{\bibinfo{volume}{7}} (\bibinfo{year}{2016}).

\bibitem{giles1996optical}
\bibinfo{author}{Giles, R.} \& \bibinfo{author}{Li, T.}
\newblock \bibinfo{title}{Optical amplifiers transform long-distance lightwave
  telecommunications}.
\newblock \emph{\bibinfo{journal}{Proceedings of the IEEE}}
  \textbf{\bibinfo{volume}{84}}, \bibinfo{pages}{870--883}
  (\bibinfo{year}{1996}).

\bibitem{borondo2014each}
\bibinfo{author}{Borondo, J.}, \bibinfo{author}{Borondo, F.},
  \bibinfo{author}{Rodriguez-Sickert, C.} \& \bibinfo{author}{Hidalgo, C.~A.}
\newblock \bibinfo{title}{To each according to its degree: The meritocracy and
  topocracy of embedded markets}.
\newblock \emph{\bibinfo{journal}{Scientific reports}}
  \textbf{\bibinfo{volume}{4}}, \bibinfo{pages}{3784} (\bibinfo{year}{2014}).

\bibitem{cohen2010complex}
\bibinfo{author}{Cohen, R.} \& \bibinfo{author}{Havlin, S.}
\newblock \emph{\bibinfo{title}{Complex networks: structure, robustness and
  function}} (\bibinfo{publisher}{Cambridge University Press},
  \bibinfo{year}{2010}).

\bibitem{kleinberg2000navigation}
\bibinfo{author}{Kleinberg, J.~M.}
\newblock \bibinfo{title}{Navigation in a small world}.
\newblock \emph{\bibinfo{journal}{Nature}} \textbf{\bibinfo{volume}{406}},
  \bibinfo{pages}{845--845} (\bibinfo{year}{2000}).

\bibitem{gulyas2015navigable}
\bibinfo{author}{Guly{\'a}s, A.}, \bibinfo{author}{B{\'\i}r{\'o}, J.~J.},
  \bibinfo{author}{K{\H{o}}r{\"o}si, A.}, \bibinfo{author}{R{\'e}tv{\'a}ri, G.}
  \& \bibinfo{author}{Krioukov, D.}
\newblock \bibinfo{title}{Navigable networks as nash equilibria of navigation
  games}.
\newblock \emph{\bibinfo{journal}{Nature communications}}
  \textbf{\bibinfo{volume}{6}} (\bibinfo{year}{2015}).

\bibitem{papadopoulos2012popularity}
\bibinfo{author}{Papadopoulos, F.}, \bibinfo{author}{Kitsak, M.},
  \bibinfo{author}{Serrano, M.~{\'A}.}, \bibinfo{author}{Bogun{\'a}, M.} \&
  \bibinfo{author}{Krioukov, D.}
\newblock \bibinfo{title}{Popularity versus similarity in growing networks}.
\newblock \emph{\bibinfo{journal}{Nature}} \textbf{\bibinfo{volume}{489}},
  \bibinfo{pages}{537--540} (\bibinfo{year}{2012}).

\bibitem{malkov2015growing}
\bibinfo{author}{Malkov, Y.~A.} \& \bibinfo{author}{Ponomarenko, A.}
\newblock \bibinfo{title}{Growing homophilic networks are natural navigable
  small worlds}.
\newblock \emph{\bibinfo{journal}{Plos ONE}} \bibinfo{pages}{e0158162}
  (\bibinfo{year}{2016}).

\bibitem{serrano2008self}
\bibinfo{author}{Serrano, M.~A.}, \bibinfo{author}{Krioukov, D.} \&
  \bibinfo{author}{Bogun{\'a}, M.}
\newblock \bibinfo{title}{Self-similarity of complex networks and hidden metric
  spaces}.
\newblock \emph{\bibinfo{journal}{Physical review letters}}
  \textbf{\bibinfo{volume}{100}}, \bibinfo{pages}{078701}
  (\bibinfo{year}{2008}).

\bibitem{boguna2009navigability}
\bibinfo{author}{Boguna, M.}, \bibinfo{author}{Krioukov, D.} \&
  \bibinfo{author}{Claffy, K.~C.}
\newblock \bibinfo{title}{Navigability of complex networks}.
\newblock \emph{\bibinfo{journal}{Nature Physics}}
  \textbf{\bibinfo{volume}{5}}, \bibinfo{pages}{74--80} (\bibinfo{year}{2009}).

\bibitem{krioukov2010hyperbolic}
\bibinfo{author}{Krioukov, D.}, \bibinfo{author}{Papadopoulos, F.},
  \bibinfo{author}{Kitsak, M.}, \bibinfo{author}{Vahdat, A.} \&
  \bibinfo{author}{Bogun{\'a}, M.}
\newblock \bibinfo{title}{Hyperbolic geometry of complex networks}.
\newblock \emph{\bibinfo{journal}{Physical Review E}}
  \textbf{\bibinfo{volume}{82}}, \bibinfo{pages}{036106}
  (\bibinfo{year}{2010}).

\bibitem{allard2017geometric}
\bibinfo{author}{Allard, A.}, \bibinfo{author}{Serrano, M.~{\'A}.},
  \bibinfo{author}{Garc{\'\i}a-P{\'e}rez, G.} \&
  \bibinfo{author}{Bogu{\~n}{\'a}, M.}
\newblock \bibinfo{title}{The geometric nature of weights in real complex
  networks}.
\newblock \emph{\bibinfo{journal}{Nature Communications}}
  \textbf{\bibinfo{volume}{8}}, \bibinfo{pages}{14103} (\bibinfo{year}{2017}).

\bibitem{barabasi1999emergence}
\bibinfo{author}{Barab{\'a}si, A.-L.} \& \bibinfo{author}{Albert, R.}
\newblock \bibinfo{title}{Emergence of scaling in random networks}.
\newblock \emph{\bibinfo{journal}{Science}} \textbf{\bibinfo{volume}{286}},
  \bibinfo{pages}{509--512} (\bibinfo{year}{1999}).

\bibitem{xenarios2002dip}
\bibinfo{author}{Xenarios, I.} \emph{et~al.}
\newblock \bibinfo{title}{Dip, the database of interacting proteins: a research
  tool for studying cellular networks of protein interactions}.
\newblock \emph{\bibinfo{journal}{Nucleic acids research}}
  \textbf{\bibinfo{volume}{30}}, \bibinfo{pages}{303--305}
  (\bibinfo{year}{2002}).

\bibitem{wiegmans_2016_47317}
\bibinfo{author}{Wiegmans, B.}
\newblock \bibinfo{title}{Gridkit: European and north-american extracts}
  (\bibinfo{year}{2016}).
\newblock \urlprefix\url{https://doi.org/10.5281/zenodo.47317}.

\end{thebibliography}

\section*{Acknowledgements}

The research work leading to these results was partially supported by HSNLab and Ericsson. Project no. 108947 and 123957 has been implemented with the support provided from the National Research, Development and Innovation Fund of Hungary, financed under the OTKA-K (2013/1) funding scheme.
Yury Malkov is grateful for the support from RFBR, according to the research project No. 16-31-60104 mol\_a\_dk and by the Government of Russian Federation (agreement \#14.Z50.31.0033 with the Institute of Applied Physics of RAS).

\section*{Author contributions statement}

M.Cs. and A.G. have developed the model and contributed to the experiments.  A.K., J.B., Z.H. and Y.M. contributed to the analysis and the numerical results. All authors reviewed the manuscript.

\section*{Additional information}

% To include, in this order: \textbf{Accession codes} (where applicable); 

\textbf{Competing financial interests}

The authors declare no competing financial interests.

% The corresponding author is responsible for submitting a \href{http://www.nature.com/srep/policies/index.html#competing}{competing financial interests statement} on behalf of all authors of the paper. This statement must be included in the submitted article file.

\end{document}